\documentclass[11pt]{article}

\usepackage[T1]{fontenc}
\usepackage{mathptmx}
\usepackage{fancyhdr}
\usepackage{graphicx}
\usepackage{cite}
\usepackage{subfigure}
\usepackage{amsmath}
\usepackage{float}
\usepackage{stfloats}
\usepackage{algorithmic}
\usepackage{algorithm}
\usepackage{amssymb}

\begin{document}

\title{\huge Optimal Power Control for Concurrent Transmissions of Location-aware Mobile Cognitive Radio Ad Hoc Networks}
\author{Yi Song and Jiang (Linda) Xie  \\            
Department of Electrical and Computer Engineering \\
The University of North Carolina at Charlotte \\
Email: \{ysong13, jxie1\}@uncc.edu}
\date{}
\maketitle

\begin{abstract}
In a cognitive radio (CR) network, CR users intend to operate over the same spectrum band licensed to legacy networks. A tradeoff exists between protecting the communications in legacy networks and maximizing the throughput of CR transmissions, especially when CR links are unstable due to the mobility of CR users. Because of the non-zero probability of false detection and implementation complexity of spectrum sensing, in this paper, we investigate a sensing-free spectrum sharing scenario for mobile CR ad hoc networks to improve the frequency reuse by incorporating the location awareness capability in CR networks. We propose an optimal power control algorithm for the CR transmitter to maximize the concurrent transmission region of CR users especially in mobile scenarios. Under the proposed power control algorithm, the mobile CR network achieves maximized throughput without causing harmful interference to primary users in the legacy network. Simulation results show that the proposed optimal power control algorithm outperforms the algorithm with the fixed power policy in terms of increasing the packet delivery ratio in the network.

\end{abstract}

\section{Introduction}
\label{introduction}
Current wireless networks are featured by the fixed spectrum assignment policy, while according to Federal Communications Commission (FCC), such fixed spectrum policy results in inefficient spectrum usage \cite{FCC-2003}. Cognitive radio (CR) technology is one of the key technologies to solve this problem by allowing opportunistic spectrum access \cite{Mitola00}. Therefore, CR networks, equipped with spectrum-aware capability, will vastly increase the spectrum utilization efficiency.

Many challenges exist in the deployment of CR networks \cite{Akyildiz-Lee06}. First of all, the transmission of CR users should not cause interference to primary (PR) users. Secondly, the throughput of CR links should be maximized for reliable quality communications. Thirdly, the robustness of CR links becomes extremely difficult to achieve under the mobility of CR users. A number of studies have been conducted in order to address these challenges.

One commonly known technique to address the above challenges is spectrum sensing, under which a CR transmitter can access the frequency band of interest only if the PR transmission is detected to be off. Through spectrum sensing, CR users can exploit unused spectrum opportunistically in a radio environment. Several spectrum detection techniques have been proposed, such as the detection of a primary transmitter through matched filter detection, energy detection, and cyclostationary feature detection \cite{Fehske05}, and the detection of local oscillator power \cite{Wild05}.

In this paper, we consider to achieve the above mentioned goals from a different perspective. Due to the non-zero probability of false detection and implementation complexity of spectrum sensing, we may raise a question: is there a way to achieve the goals of CR networks without spectrum sensing? Hence, we study a new sensing-free solution to enable concurrent transmissions of mobile CR users and also guarantee non-interference to PR users, thus improve the frequency reuse. With such aim, we examine a location-aware spectrum sharing scenario, where a CR ad hoc network is overlaid to a legacy network. CR users intend to operate over the same spectrum band which is licensed to PR users. The objective is to maximize the concurrent transmission region of CR users within which they can move, while at the same time maintaining non-interference to PR communications. 
\begin{figure}[htb!]
\centering
\subfigure[Without power control]
{\includegraphics[scale=0.45,clip]{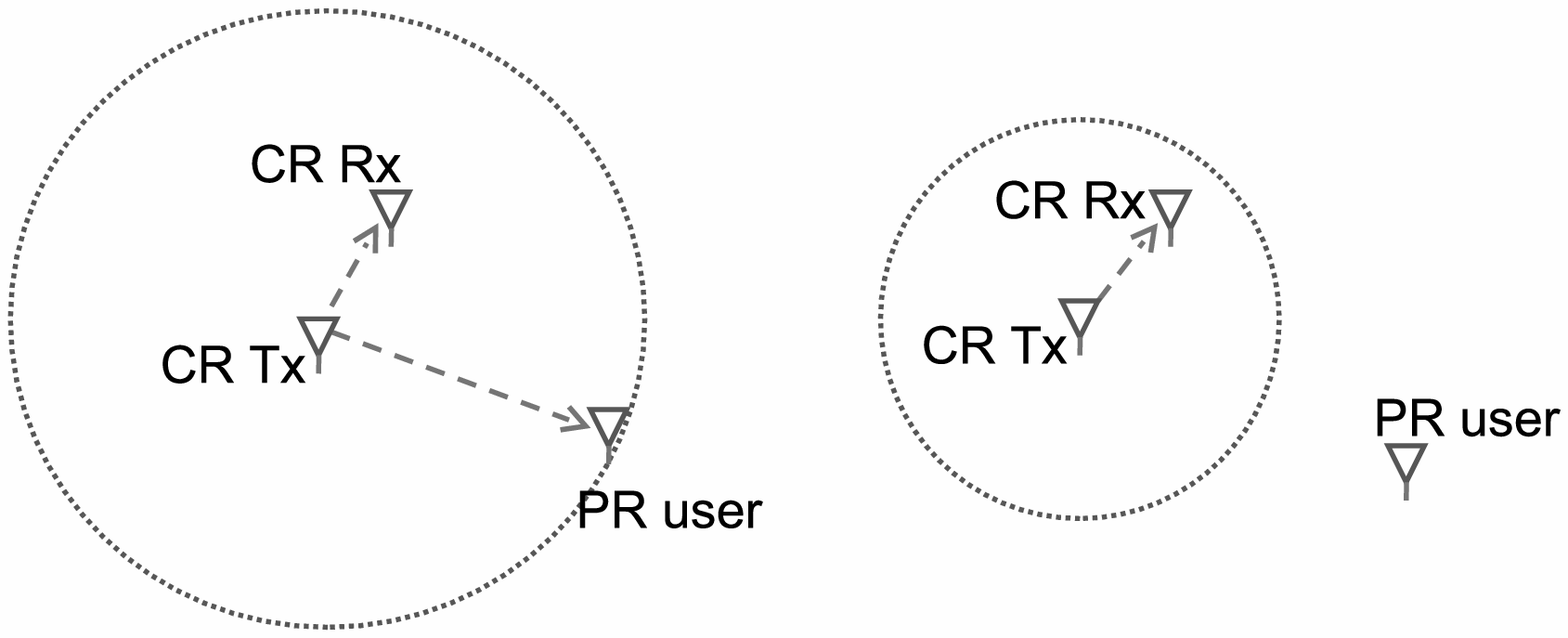}\label{fig:withoutpowercontrol}}
\subfigure[With power control]
{\includegraphics[scale=0.45,clip]{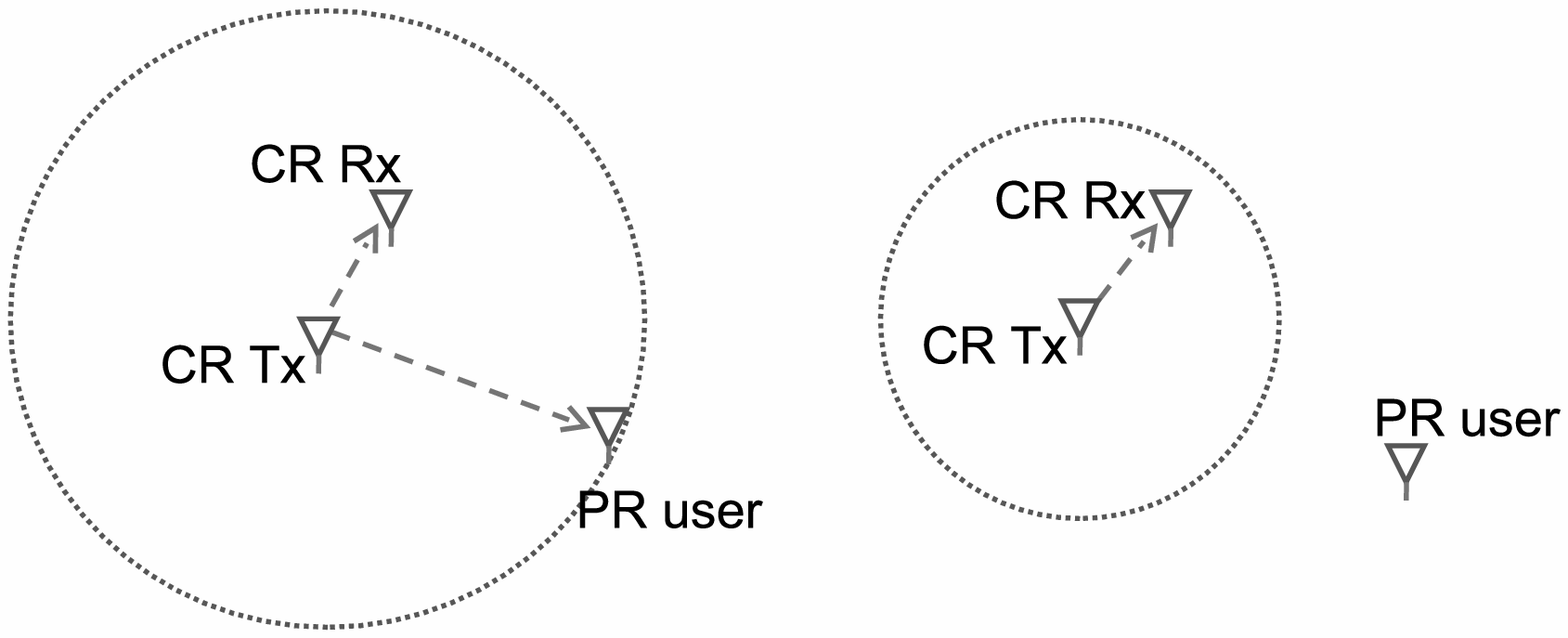}\label{fig:withpowercontrol}}
\vspace{-0.05in}
\caption{A spectrum sharing scenario of a two-node CR ad hoc network with a PR user.}
\label{fig:powercontrol}
\vspace{-0.05in}
\end{figure}

To achieve the above objective, power control policies are important to guarantee the quality of both CR and PR communications. Fig. \ref{fig:powercontrol} demonstrates the scenario of using a fixed power policy (in (a)) and the scenario of using power control (in (b)). The figure indicates that without power control, when a PR user is within the interference range of a CR transmission, concurrent transmissions are not possible. However, with power control, concurrent transmissions become feasible by reducing the transmit power of the CR transmitter to ensure non-interference to the PR user. Hence, the concurrent transmission region defined in this paper refers to the circle within which the transmissions of CR users can be conducted without interfering PR users. The optimal power defined in this paper refers to the transmit power which makes the concurrent transmission region of a CR user the maximum so that the bandwidth efficiency and CR link throughput can be improved. In addition, we assume that every node has its own location information in the system through Global Positioning System (GPS) or other positioning algorithms \cite{Celebi07}, and every node is able to exchange location information via a common control channel with its neighboring nodes \cite{Hur09}\cite{Zhang02}.

Currently, related work on power control and concurrent transmissions of CR networks falls into two categories. In the first category, the power control problem is considered in terms of either improving network energy efficiency \cite{qian07,LiChinacom08,sujsac08,Tang08}, or supporting user communication sessions in multi-hop CR networks \cite{Shiinfocom07}, but the concurrent transmission for CR users is not considered. On the other hand, in \cite{Wang09-C}, the scanning-free concurrent transmission region for CR users is considered only from a geometric point of view without taking power control into account. In addition, in this work, the CR transmitters and receivers are geographically fixed and the mobility of CR users is ignored. The concurrent transmission area defined in \cite{Wang09-C} is an irregular area which is difficult to apply in mobile scenarios. In \cite{Hur09}, a location-assisted MAC protocol is proposed to enable concurrent transmissions for exposed nodes. In \cite{Hov05}, the power scaling constraint of a CR transmitter is studied.

Our proposed optimal power control algorithm differs from related work in the original motivations. Most related work only considers fixed transmit power at each CR node without the power control capability \cite{Wang09-C} \cite{Behzad05}. In this paper, we study a mobile CR network where each CR node has the power control capability. That is, each CR node can transmit at any power in the allowable transmit power range to achieve the maximum concurrent transmission region.  Our main contribution is that we propose a location-aware sensing-free optimal power control algorithm for concurrent transmissions especially in mobile CR ad hoc networks. Under such algorithm, the CR transmitter is able to conduct transmissions with the presence of the PR users while moving. Even if the CR users are in the area called ``protected region'' \cite{Hov05} in which the CR users should not transmit, if the location information of both CR and PR receivers is known to the CR transmitter, the CR transmitter can adjust its transmit power to enable the concurrent transmission.

The rest of this paper is organized as follows. In Section \ref{sc:system model}, the system model of a CR ad hoc network overlaid to a legacy network is described. The formulation of the optimal power control problem is also given. In Section \ref{sc:optimal power control}, the proposed optimal power control algorithm for concurrent transmissions of CR networks is explained. Simulation results are presented in Section \ref{sc:simulation}, followed by conclusions in Section \ref{sc:conclusion}.

\section{System Model and Problem Formulation}
\label{sc:system model}
\begin{figure}[t]
\centerline{\includegraphics[scale=0.44,clip]{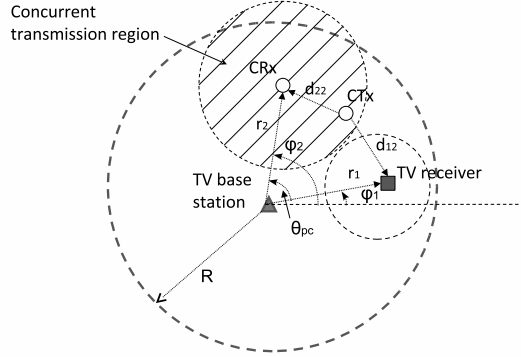}}
\vspace{-0.09in}
\caption{System model of a CR ad hoc network overlaid to a legacy network.}
\label{fig:model}
\vspace{-0.15in}
\end{figure}

In this section, a spectrum sharing scenario in which a cognitive radio ad hoc network overlaid to a legacy network is considered. Fig. \ref{fig:model} shows the system model, where the shaded triangle and square represent the PR transmitter and receiver, respectively. The white circles are the CR transmitter (denoted as CTx) and receiver (denoted as CRx). They form an ad hoc network to share the same spectrum band with the primary network. Without loss of generality, we assume that the PR base station is at the origin of the coordinate axes, and the PR receiver does not move. Let the location of the PR and CR receiver be $(r_{1}, \varphi_{1})$ and $(r_{2},\varphi_{2})$, respectively. $d_{12}$ represents the distance between the CTx and the PR receiver, and $d_{22}$ represents the distance between the CTx and CRx. The decodable radius of the TV base station is $R$. Thus, the distance between the PR and CR receivers is 
$d_{pc} = \sqrt{r^2_{1} + r^2_{2} - 2r_1r_2\cos\theta_{pc}}$, where $\theta_{pc}$ is the relative angle of the PR and CR receivers.

Based on the two-ray ground propagation model \cite{Rappaport}\cite{Tang07}, the received signal power $P_{r}$ can be written as$
P_r= \frac{P_tG_tG_rh^2_th^2_r}{r^\alpha}$, where $P_t$ is the transmit power, $G_t$ and $G_r$ are the gains of the transmitter and receiver antennas, respectively; $h_t$ and $h_r$ are the heights of the transmitter and receiver antennas, respectively; $r$ is the distance between the transmitter and the receiver; and $\alpha$ is the path loss factor. 

In this paper, we consider that the concurrent transmission for CR users must satisfy the co-channel signal-to-interference ratio (SIR) requirements for both PR and CR receivers. We denote the SIR thresholds for the PR and CR receivers as $\tau_p$ and $\tau_c$, respectively; and the SIRs for the PR and CR receivers are $SIR_p$ and $SIR_c$, respectively. The optimal power control problem for concurrent transmission region maximization is formulated as follows:\\
Maximize: \hspace{0.02in} the area of concurrent transmission region \\
Subject to: \hspace{0.02in}
\vspace{-0.1in}
\begin{equation}\label{eq:second}
\begin{split}
SIR_p > \tau_p \\
SIR_c > \tau_c \\
P^{min}_c \le P_{ct} \le P^{max}_c,
\end{split}
\end{equation}
where $P_{ct}$ is the transmit power of the CTx, $P^{min}_c$ and $P^{max}_c$ are the minimum and maximum allowable transmit power of the CTx, respectively.

\section{Optimal Power Control}
\label{sc:optimal power control}
In this section, the proposed optimal power control algorithm for concurrent transmission region maximization is presented. We first consider the feasibility of the proposed optimal power control algorithm. Then, we consider the implementation of the algorithm in a mobility scenario. Finally, the impact of the shadowing fading effect on the optimal power control algorithm is investigated for the mobility scenario.

\subsection{Feasibility of Optimal Power Control}
\label{ssc:feasibility}
We assume that the transmit power of the TV base station is $P_{bs}$, the gains of the transmitter and receiver antennas are unity, the heights of the antennas are the same, the path loss factors of the PR and CR transmissions are the same, and the Gaussian noise is negligible. Based on these assumptions, the SIRs at both CR and PR receivers can be written as $SIR_c = \frac{ P_{ct}r_2^\alpha}{P_{bs}d_{22}^\alpha}$ and $ SIR_p = \frac{P_{bs}d_{12}^\alpha}{P_{ct}r_1^\alpha}$, respectively. Since the SIRs must satisfy (\ref{eq:second}), we have\\
\begin{equation}\label{eq:third}
\begin{split}
d_{22} < r_2(\frac{P_{ct}}{\tau_cP_{bs}})^{1/\alpha} \\
d_{12} > r_1(\frac{\tau_pP_{ct}}{P_{bs}})^{1/\alpha}.
\end{split}
\end{equation}
The first constraint in (\ref{eq:third}) means that the CTx which can concurrently transmit to the CRx must be physically within the disk centered at the CRx with a radius of $r_2(\frac{P_{ct}}{\tau_cP_{bs}})^{1/\alpha}$, as shown in Fig. \ref{fig:model}. The second constraint means that the CTx must not fall into the disk which is centered at the TV receiver with a radius of $r_1(\frac{\tau_pP_{ct}}{P_{bs}})^{1/\alpha}$, as shown in Fig. \ref{fig:model}. Therefore, the concurrent transmission region reaches the maximum when the following equation is satisfied: \\
\begin{equation}
r_1(\frac{\tau_pP_{ct}}{P_{bs}})^{1/\alpha} + r_2(\frac{P_{ct}}{\tau_cP_{bs}})^{1/\alpha} = d_{pc}. 
\label{eq:optimal}
\end{equation}
Hence, given $r_1$, $r_2$, and $\theta_{pc}$, the optimal power for concurrent transmission region maximization can be derived by solving equation (\ref{eq:optimal}).

\begin{figure}[t]
\centering
\subfigure[]{\includegraphics[scale=0.32,clip]{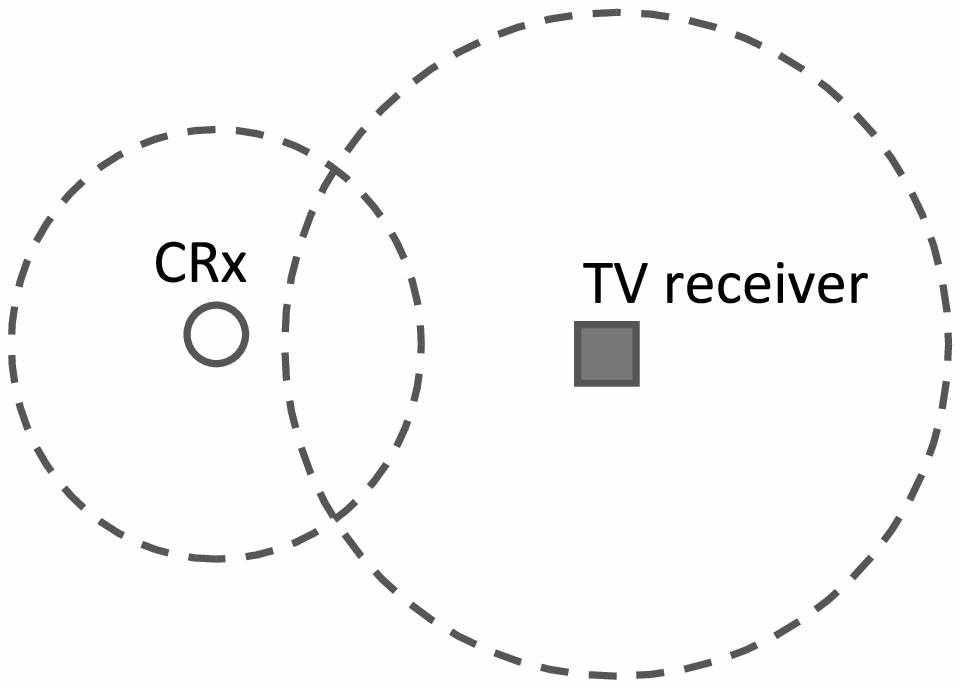}\label{fig:sub1}}
\subfigure[]{\includegraphics[scale=0.32,clip]{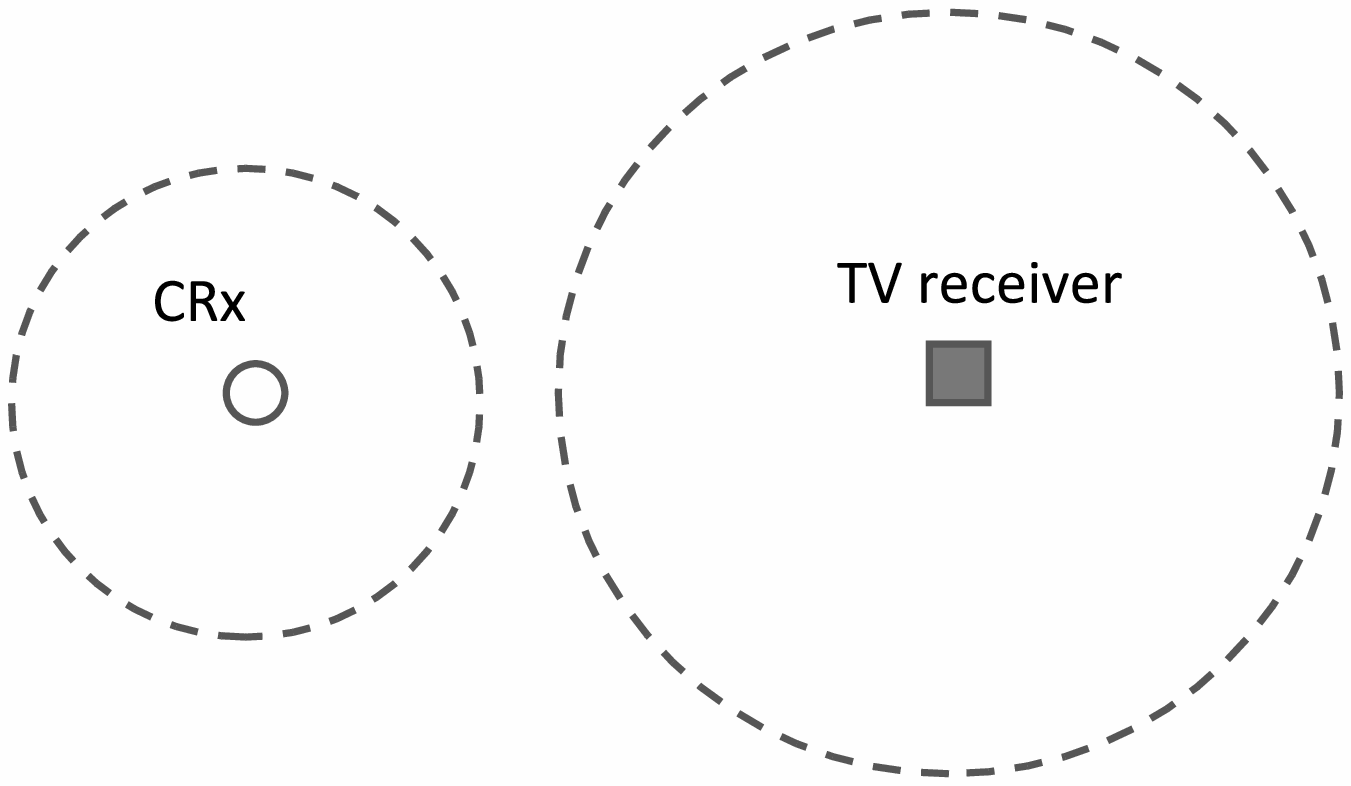}\label{fig:sub2}}
\caption{Two possible cases that there will be no solution for (\ref{eq:optimal}). (a) $f(r_2,\theta_{pc})>0$. (b) $g(r_2,\theta_{pc})<0$.}
\label{fig:twocases}
\end{figure}

However, considering (\ref{eq:second}), the solution of (\ref{eq:optimal}) may not lie in the allowable range [$P_c^{min}$, $P_c^{max}$]. So we consider two extreme cases by letting $P_{ct}$ be $P^{min}_c$ and $P^{max}_c$, respectively. We have the following two extreme functions of $r_2$ and $\theta_{pc}$:\\
\begin{equation}
\begin{split}
f(r_2,\theta_{pc}) = & r_1(\frac{\tau_pP^{min}_c}{P_{bs}})^{1/\alpha} + r_2(\frac{P^{min}_c}{\tau_cP_{bs}})^{1/\alpha} \\
& - \sqrt{r^2_{1} + r^2_{2} - 2r_1r_2\cos\theta_{pc}}
\end{split}
\label{eq:lower bound}
\end{equation}
\begin{equation}
\begin{split}
g(r_2,\theta_{pc}) = &r_1(\frac{\tau_pP^{max}_c}{P_{bs}})^{1/\alpha} + r_2(\frac{P^{max}_c}{\tau_cP_{bs}})^{1/\alpha} \\
&- \sqrt{r^2_{1} + r^2_{2} - 2r_1r_2\cos\theta_{pc}}.
\end{split}
\label{eq:upper bound}
\end{equation}
If $f(r_2,\theta_{pc})>0$, as shown in Fig. \ref{fig:sub1}, the two disks overlap and increasing the transmit power $P_{ct}$ cannot make these two disks separate, therefore, the optimal transmit power of $P_{ct}$ can never be reached. Similarly, if $g(r_2,\theta_{pc})<0$, as shown in Fig. \ref{fig:sub2}, there will not be an optimal power either. Hence, the existence of the optimal $P_{ct}$ power relies on $r_2$ and $\theta_{pc}$. If the optimal power control is feasible, the two cases shown in Fig. \ref{fig:twocases} should be avoided. That is, to let the optimal power exist, $r_2$ and $\theta_{pc}$ must be in the set\\
\begin{equation}
\{(r_2,\theta_{pc})|f(r_2,\theta_{pc})\le 0 \cap g(r_2,\theta_{pc}) \geq 0)\}.
\label{eq:set}
\end{equation}

\begin{figure}[htb!]
\centering
\small
\begin{tabular}{l}
\hline
update $r_1$ , $\varphi_1$, $r_2$ and $\varphi_2$;  \\
calculate $\theta_{pc}$, $d_{22}$, $f(r_2,\theta_{pc})$ and $g(r_2,\theta_{pc})$; \\
\textbf{if} $(f(r_2,\theta_{pc})\le 0)$AND$(g(r_2,\theta_{pc}) \geq 0)$AND$(d_{22} \le r_{max})$ \\
\hspace{0.05in} calculate optimal power;   
\hspace{0.15in}//optimal power could apply  \\
\hspace{0.05in} transmit with optimal power; \\
\textbf{elseif} $(g(r_2,\theta_{pc})<0)$AND$(d_{22} \le r_{max})$    \\
\hspace{0.05in} transmit with maximum power; 
//concurrent transmission\\
\hspace{1.75in} will not affect primary user \\
\textbf{elseif} $(f(r_2,\theta_{pc})>0)$ \\
\hspace{0.05in} stop transmitting;   
\hspace{0.4in}//concurrent transmission is not \\
\hspace{1.5in} allowed \\
\textbf{endif} \\
\hline
\end{tabular}
\label{fig:fix algo}
\caption{Power control algorithm for fixed CRx.}
\vspace{-0.14in}
\end{figure}
Fig. 4 presents the proposed optimal power control algorithm for the scenario when the CRx is in a fixed location, where $r_{max}$ is the maximum decodable range of the CTx. Recall that the location information of both the CR and PR receivers is available to the CTx. The proposed algorithm first evaluates the feasibility of the optimal power control for concurrent transmissions. Then, the optimal power can be computed using any numerical method when the optimal power control is feasible. The last case checks the availability of concurrent transmissions, and the CTx will not conduct transmissions unless the condition is violated.

\subsection{Optimal Power Control for Mobility Scenarios}
\label{ssc:mobility}
\begin{figure}[htb!]
\centerline{\includegraphics[scale=0.44,clip]{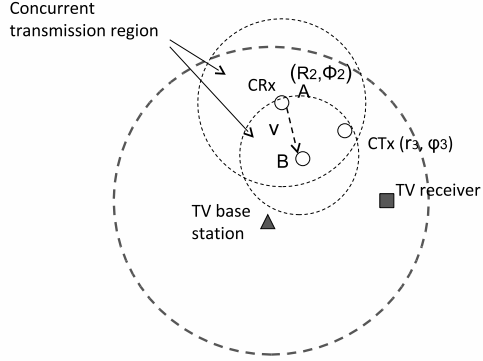}}
\caption{The concurrent transmission region evolves with the movement of the CRx.}
\label{fig:mobility}
\end{figure}

We now extend our model to the scenario in which the mobility of the CRx is considered. Because of the mobility of the CRx, $r_2$ and relative angle $\theta_{pc}$ change with the movement of the CRx. Thus, the optimal power and the concurrent transmission region also change. Fig. \ref{fig:mobility} shows the scenario where the concurrent transmission region evolves with the movement of the CRx from A to B.

Without loss of generality, we assume that the CR transmitter is static in a location $(r_3,\varphi_3)$, and the CRx is moving from a starting point $(R_2,\Phi_2)$ with a velocity of $\vec{v}=s\vec{u}$, where s is the speed and $\vec{u}=(\cos\gamma,\sin\gamma)$ is the unit directional vector. Therefore, the polar coordinates of the CRx can be written as:
\[
\left\{
\begin{array}{l}
	r_2(t)  =  \sqrt{(R_2\cos\Phi_2+st\cos\gamma)^2+(R_2\sin\Phi_2+st\sin\gamma)^2} \\
	\varphi_2(t)  = \arctan(\frac{R_2\sin\Phi_2+st\sin\gamma}{R_2\cos\Phi_2+st\cos\gamma}).
\end{array}
\right.
\]

If the movement pattern of the CRx does not change (i.e., the direction and velocity remain the same) or the movement pattern is deterministic, the coordinates of the CRx are just functions of time. Therefore, the CTx can ``predict'' the location of the CRx, thus adjust its transmit power using exactly the same optimal power control algorithm shown in Fig. 4.

On the other hand, if the movement pattern of the CRx keeps changing randomly, the CRx should update its location to the CTx for computing the optimal power control. Fig. 6 demonstrates the proposed optimal power control algorithm for a mobile CRx that changes movement patterns randomly, where $r_{CT}$ is the radius of the concurrent transmission region.
\begin{figure}[h]
\centering
\vspace{-0.05in}
\small
\begin{tabular}{l}
\hline
update $r_1$ , $\varphi_1$, $R_2$, $\Phi_2$, $r_3$, $\varphi_3$, $s$ and $\vec{u}$;  \\
calculate $r_2,\varphi_2$, $\theta_{pc}$ and $d_{22}$; \\
\textbf{if} ($d_{22} \le r_{CT}$)AND($d_{22} \le r_{max}$)\\
\hspace{1in}//the distance between CTx and CRx \\
\hspace{1in} is still in concurrent transmission region \\
\hspace{0.07in} transmit power remains the same; \\
\textbf{else} \\
\hspace{0.05in} calculate $f(r_2,\theta_{pc})$ and $g(r_2,\theta_{pc})$;\\
\textbf{if} $(f(r_2,\theta_{pc})\le 0$)AND($g(r_2,\theta_{pc}) \geq 0)$AND$(d_{22} \le r_{max})$ \\
\hspace{0.05in} calculate optimal power; \\
\hspace{0.05in} calculate concurrent transmission radius $r_{CT}$; \\
\hspace{0.05in} transmit with optimal power; \\
\textbf{elseif} $(g(r_2,\theta_{pc})<0)$AND$(d_{22} \le r_{max})$    \\
\hspace{0.05in} transmit with maximum power;
//concurrent transmission \\
\hspace{1.7in} will not affect primary user \\
\textbf{elseif} $(f(r_2,\theta_{pc})>0)$ \\
\hspace{0.05in} stop transmitting;
\hspace{0.4in}//concurrent transmission is not \\
\hspace{1.5in} allowed \\
\textbf{endif} \\
\textbf{endif} \\
\hline
\end{tabular}
\label{fig:mobalgo}
\caption{Power control algorithm for the mobile CRx.}
\vspace{-0.05in}
\end{figure}

\subsection{Shadowing Fading Effect}
\label{ssc:shadowing effect}
In this subsection, we consider the impact of the shadowing fading effect on the optimal power control algorithm. Since the antenna of the TV transmitter is usually hundreds of meters higher than that of the CR transmitter, we loose the assumption that the path loss factors of the PR user and CR user are the same, and assume $\alpha_1<\alpha_2$, where $\alpha_1$ and $\alpha_2$ are the path loss factors of the PR user and CR user, respectively. Using log-distance path loss model \cite{Rappaport}, the path loss of PR transmissions can be written as:
	\[
	PL_p(r_1)[dB]=PL_p(d_0)+10\alpha_1\log(\frac{r_1}{d_0})+X_\sigma,
\]
where $d_0$ is the reference distance and $X_\sigma$ is a zero-mean Gaussian random variable with standard deviation $\sigma$ which is location and distance dependent. Therefore, the received power of the PR receiver is $P_{pr}(r_1)=P_{bs}-PL_p(r_1)$, and interference from the CTx is $P_i(d_{12})=P_{ct}-PL_c(d_{12})$. Hence, the SIR at the PR receiver is $ SIR_p=P_{pr}(r_1)-P_i(d_{12})$. Similarly, the SIR at the CR receiver is  $SIR_c=P_{cr}(d_{22})-P_i(r_2)$. Since the SIRs must satisfy the constraints in (\ref{eq:second}), we have
\begin{equation}
d_{12}[dB]> \frac{P_{ct}+\alpha_1r_1+\tau_p+X_\sigma^{'}-P_{bs}}{\alpha_2}
\label{eq:eight}
\end{equation}
\begin{equation}
d_{22}[dB]< \frac{P_{ct}+\alpha_1r_2-\tau_c-X_\sigma^{'}-P_{bs}}{\alpha_2},
\label{eq:nine}
\end{equation}
where $X_\sigma^{'}\sim N(0,\sqrt{2}\sigma)$. Similar to (\ref{eq:optimal}), the optimal power is achieved when the following equation is satisfied.
\begin{equation}
\begin{split}
& 10^{\frac{P_{ct}+\alpha_1r_1+\tau_p+X_\sigma^{'}-P_{bs}}{10\alpha_2}}+10^{\frac{P_{ct}+\alpha_1r_2-\tau_c-X_\sigma^{'}-P_{bs}}{10\alpha_2}}\\
&=\sqrt{r_1^2+r_2^2-2r_1r_2\cos\theta_{pc}}.
\end{split}
\label{eq:ten}
\end{equation}
The solution of equation (\ref{eq:ten}) can be written as 
\begin{equation}
P_{ct}[dB] =10\alpha_2\lg\left( \frac{\sqrt{r_1^2+r_2^2-2r_1r_2\cos\theta_{pc}}}{10^{\frac{\alpha_1r_1+\tau_p+X_\sigma^{'}-P_{bs}}{10\alpha_2}}+10^{\frac{\alpha_1r_2-\tau_c-X_\sigma^{'}-P_{bs}}{10\alpha_2}}}\right).
\label{eq:ten1}
\end{equation}
%

\section{Performance Results}
\label{sc:simulation}
In this section, the performance of the proposed optimal power control algorithm is evaluated via simulations and compared with the power control algorithm with fixed transmit power.

\subsection{Simulation Parameters}
\label{ssc:parameters}
The parameters used in our simulations are listed in Table \ref{tb:parameters}. We assume that the transmit power of the TV base station is 100 kW \cite{FCC-2004-186}, the transmit power range of the CTx is [1W, 100W] \cite{FCC-2004-186}, and the SIR thresholds for the TV and CR receivers are 30dB and 3dB, respectively.
\begin{table}[h]\caption{Simulation Parameters}
\normalsize
\centering
\begin{tabular}{|l|c|}\hline
	TV base station transmit power & 100kW \\ \hline
	maximum transmit power of CTx & 100W \\ \hline
	minimum transmit power of CTx & 1W \\ \hline
  coordinates of TV receiver & $(50km, 0^{\circ})$ \\ \hline
  coordinates of CTx & $(50km, 60^{\circ})$ \\ \hline
  SIR thershold for PR receiver & 30dB \\ \hline
  SIR thershold for CR receiver & 3dB \\ \hline
  path loss factor & 3 \\ \hline
  Simulation time & 1000s \\
	\hline
\end{tabular}
\label{tb:parameters}
\end{table}

The mobility characteristics of the CRx are modeled using the random waypoint mobility model \cite{Camp02}\cite{Zhang06}. The CRx changes its movement pattern every $ts$ seconds, where $ts$ is uniformly distributed between 0 and 30s. The average speed of the CRx $s$ is chosen at 10, 20, 30, 40 m/s. The heading angle of the CRx is selected to be uniformly distributed between 0 and $2\pi$. The average pause time of the CRx is set to be 5 seconds. The starting position of the CRx is $(50km, \ 60^{\circ})$. The time-based update mechanism is used in our simulations with the time threshold 1 second.

The length of packets sent from the CTx is exponentially distributed with the mean length of 100 bytes. The packets are sent in a Poisson stream fashion with the average arrival rate of 10 packets/s.

\subsection{Simulation Results}
\label{ssc:results}
\begin{figure}[htb!]
\centering
\includegraphics[scale=0.5]{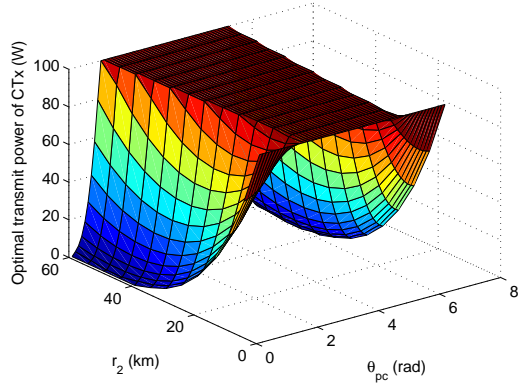}
	\caption{Optimal transmit power of CTx for maximum concurrent transmission.}
	\label{fig:optpwr}
\end{figure}
First, from (\ref{eq:optimal}) we obtain the plane of the optimal power with respect to $r_2$ and $\theta_{pc}$, as shown in Fig. \ref{fig:optpwr}. It is observed that when $\theta_{pc}$ is within a certain range, the optimal power is constant at $P_c^{max}$. This is because that if the solution of (\ref{eq:optimal}) is greater than the maximum allowable transmit power of the CTx, the optimal power will be limited to the maximum transmit power.
\begin{figure}[htb!]
	\centering
		\includegraphics[scale=0.5]{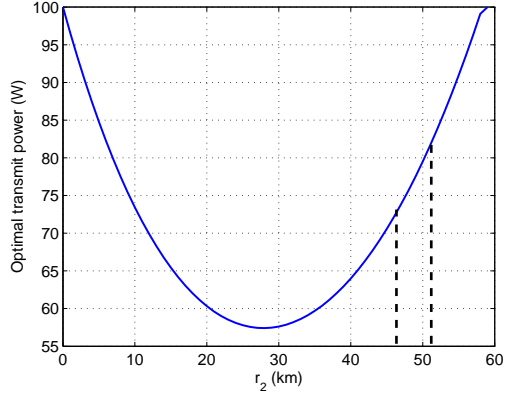}
		\caption{Optimal transmit power when $\theta_{pc}=60^{\circ}$.}
	\label{fig:powervsr2}
\end{figure}
\begin{figure}[htb!]
	\centering
	\includegraphics[scale=0.5]{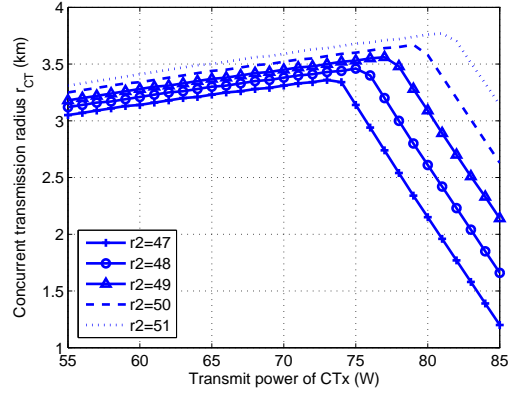}
			\caption{Simulation results of concurrent transmission radius vs. transmit power of CTx.}
	\label{fig:powervsradius}
\end{figure}

Fig. \ref{fig:powervsr2} along with Fig. \ref{fig:powervsradius} illustrate the relationship between the radius of the concurrent transmission region $r_{CT}$ and the transmit power of the CTx under different $r_2$. Fig. \ref{fig:powervsr2} is obtained from Fig. \ref{fig:optpwr} when $\theta_{pc}$ is fixed to be $60^{\circ}$, while Fig. \ref{fig:powervsradius} is obtained through the simulation based on the constraints in (\ref{eq:second}). It is noted that when $r_2$ is in the interval [47km, 51km] as shown in Fig. \ref{fig:powervsr2}, the optimal power in these two figures match perfectly, which indicate the analytical and simulation results coincided well. From the proposed optimal power control algorithm shown in Fig. 6, the distance between the CTx and the CRx $d_{22}$ must be smaller than the maximum decodable radius of the CTx to let the concurrent transmission be feasible. According to (\ref{eq:third}), the maximum decodable radii of the CTx are 3.7km when $r_2$ is 47km and 4.2km when $r_2$ is 54km. So if $r_2$ is out of the neighborhood of 50km (i.e., [47km, 54km]), $d_{22}$ is larger than the maximum decodable radius of the CTx. Hence, the concurrent transmission radius is zero, which means that the concurrent transmission is not allowed. From Fig. \ref{fig:powervsr2}, the radius of the concurrent transmission region increases as the transmit power of the CTx increases. When the transmit power reaches the optimal power, the concurrent transmission radius reaches the maximum, and then it decreases drastically.

\begin{figure}[tp]
\begin{center}
\subfigure[Average speed = 10m/s]{\includegraphics[scale=0.32]{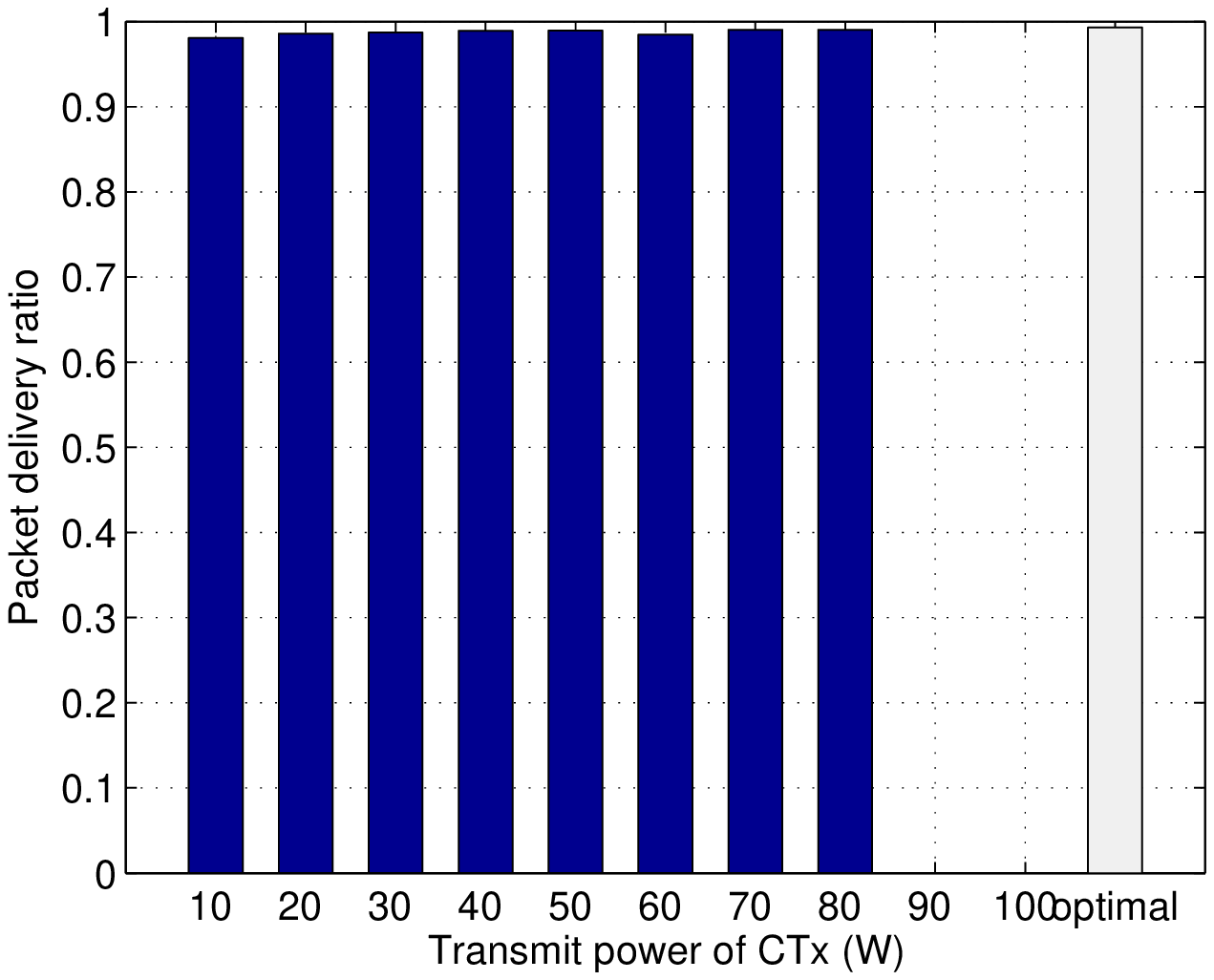}\label{fig:s10}}
\subfigure[Average speed = 20m/s]{\includegraphics[scale=0.32]{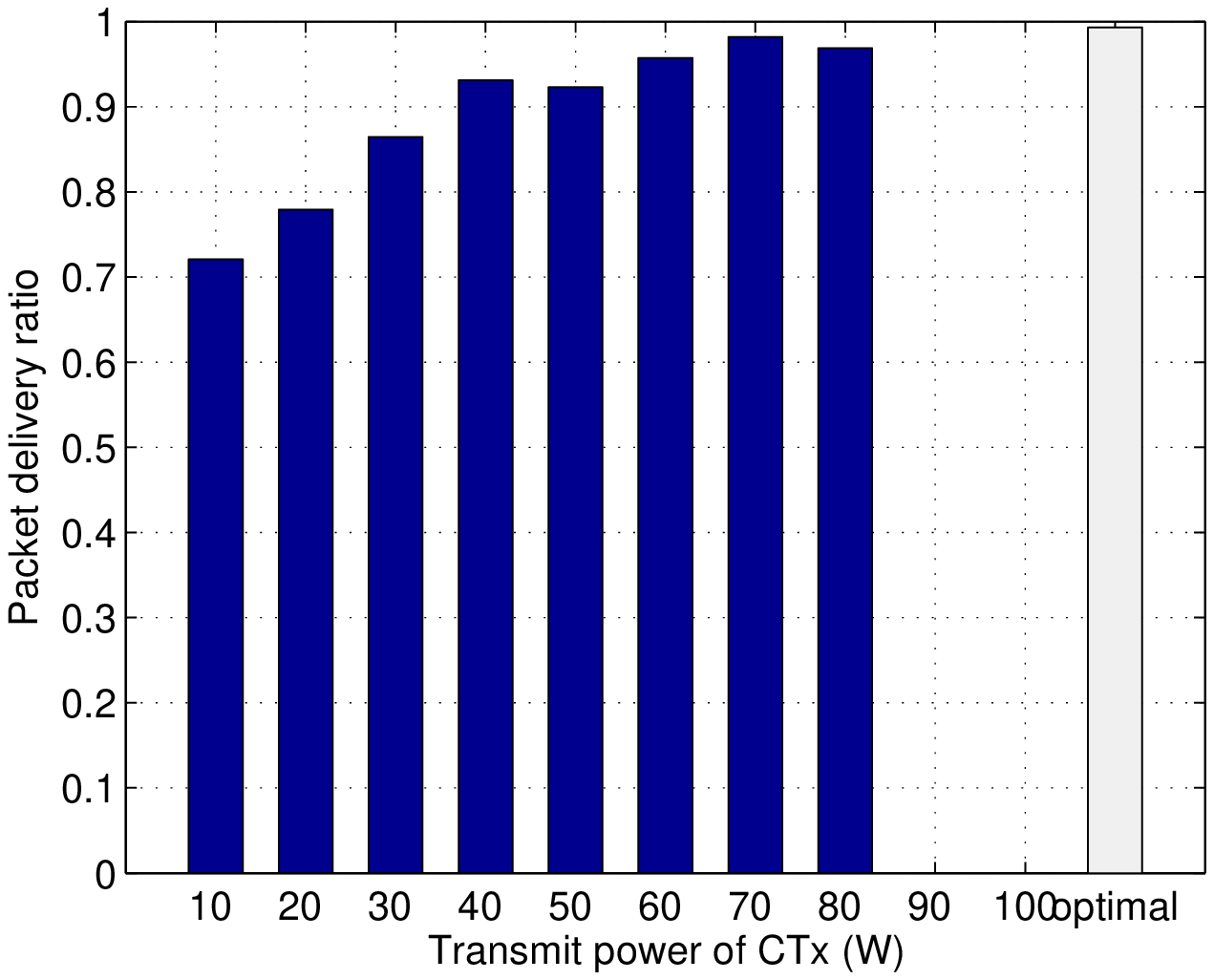}\label{fig:s20}}\\
\subfigure[Average speed = 30m/s]{\includegraphics[scale=0.32]{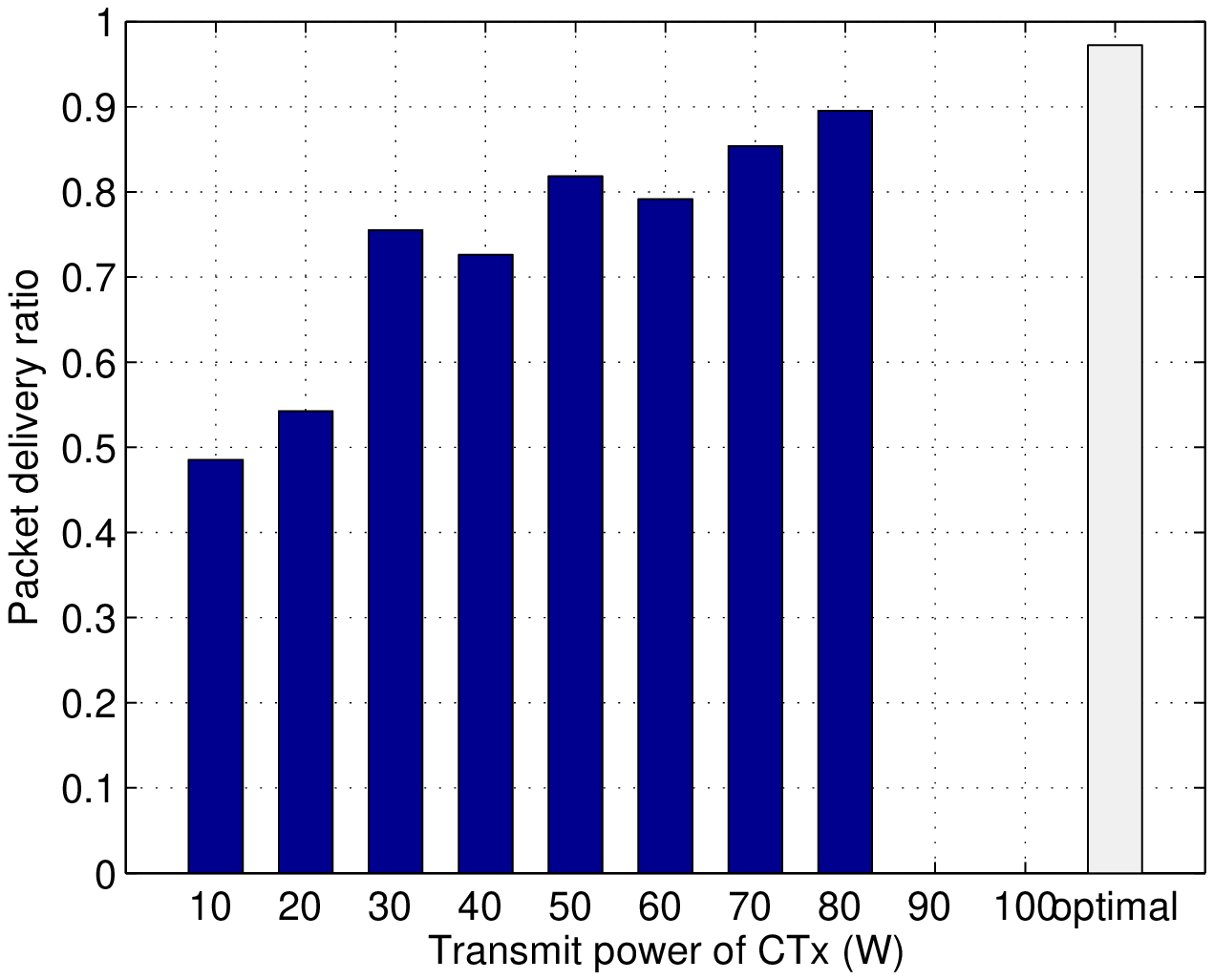}\label{fig:s30}}
\subfigure[Average speed = 40m/s]{\includegraphics[scale=0.32]{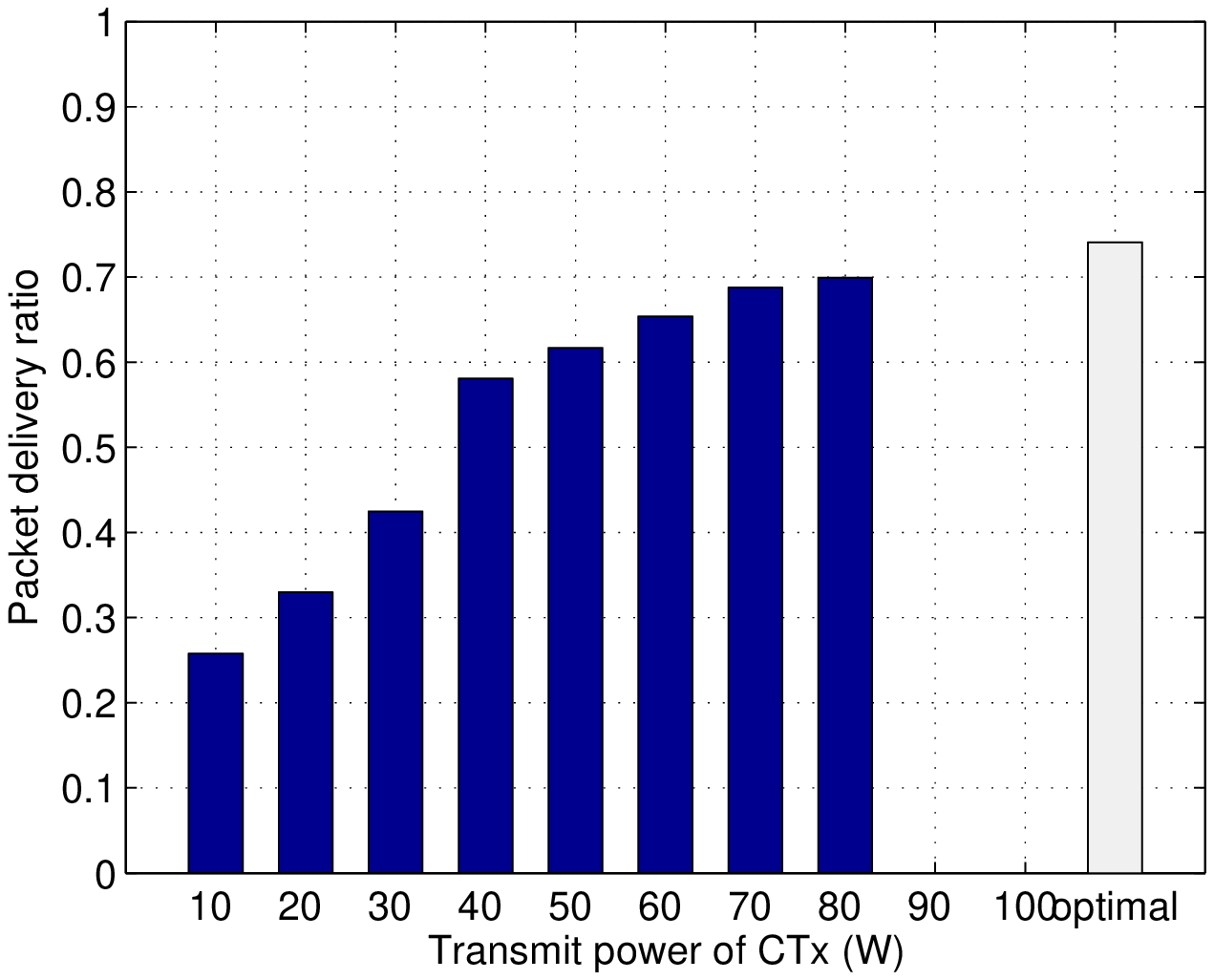}\label{fig:s40}}
\end{center}
\caption{Packet delivery ratio using fixed power algorithm and the proposed power control algorithm under different average speeds.}
\label{fig:packetratio}
\end{figure}

Fig. \ref{fig:packetratio} shows the simulation results of packet delivery ratio of the mobile CR ad hoc network using fixed transmit power of the CTx and the proposed optimal power control algorithm with different moving speeds of the CRx. The mobility characteristics are given in Section \ref{ssc:parameters}. First of all, it is observed that the overall packet delivery ratio suffers degradation as the moving speed of the CRx increases. Secondly, with the same moving speed, the packet delivery ratio increases as the transmit power of the CTx increases. When the fixed transmit power of the CTx exceeds 80W, the packet delivery ratio decreases to zero. This is because that the $SIR_p$ can never be satisfied when the transmit power of the CTx exceeds 80W. However, it is noted that the packet delivery ratio using the proposed optimal power control algorithm is always higher than that of the fixed power algorithm at any speed. 

\begin{figure}[t!]
	\centering
		\includegraphics[scale=0.5]{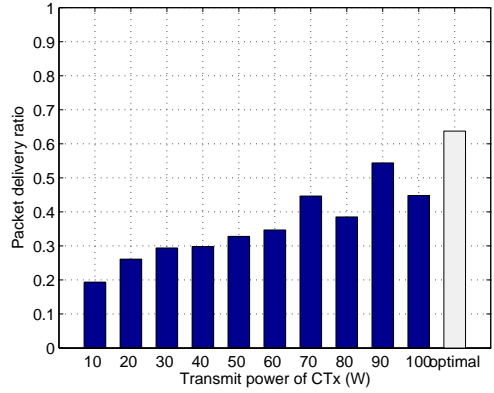}
		\caption{Packet delivery ratio under the shadowing fading effect (average speed = 30m/s).}
	\label{fig:shadowing}
\end{figure}

Finally, Fig. \ref{fig:shadowing} shows the simulation results of the packet delivery ratio under different power control algorithms with the impact of the shadowing fading effect. The mobility characteristics are the same as used for Fig. \ref{fig:packetratio}. The path loss factors of PR and CR transmissions are 3 and 4, respectively. The average speed of the CRx is set to be 30 m/s, and the standard deviation $\sigma$ is chosen to be 6 dB. Compared to Fig. \ref{fig:s30}, the overall packet delivery ratio decreases significantly. However, with the shadowing fading effect, the $SIR_p$ can be satisfied with certain probability when the transmit power of the CTx exceeds 80W. It is observed from the simulation results that the proposed optimal power control algorithm also outperforms the fixed power algorithm under the shadowing fading effect.

\section{Conclusion}
\label{sc:conclusion}
In this paper, an optimal power control algorithm for concurrent transmissions of location-aware mobile CR ad hoc networks is proposed. The proposed algorithm incorporates the mobility characteristics of the CR receiver in the algorithm design and is aimed to maximize the concurrent transmission region of CR users, hence improving the throughput of CR links. Simulation results demonstrate that the packet delivery ratio of the proposed optimal power control algorithm can be effectively improved, as compared to that of the fixed power algorithm. The impact of the shadowing fading effect on the proposed algorithm is also considered. It is shown that the proposed power control algorithm also outperforms the fixed power control algorithm under the shadowing fading effect.

\bibliographystyle{IEEEtran}
\bibliography{allreferences}

\end{document}